\documentclass[runningheads]{llncs}
\usepackage{url}
\usepackage[noadjust]{cite}
\usepackage{comment}
\usepackage{cite}
\usepackage{amsmath,amssymb,amsfonts}
\usepackage{algorithmic}
\usepackage{graphicx}
\usepackage{textcomp}
\usepackage{xcolor}
\usepackage{hyperref}

\usepackage{caption}
\usepackage{subcaption}
\usepackage{booktabs, makecell, longtable}
\usepackage{makecell}
\def\BibTeX{{\rm B\kern-.05em{\sc i\kern-.025em b}\kern-.08em
    T\kern-.1667em\lower.7ex\hbox{E}\kern-.125emX}}

\begin{document}
\title{Performance Comparison of Session-based Recommendation Algorithms based on GNNs}
\titlerunning{Performance Comparison of Session-based Recommendation Algorithms}

\author{Faisal Shehzad\orcidID{0000-0001-6239-8198}, and Dietmar Jannach\orcidID{0000-0002-4698-8507}}
\authorrunning{Shehzad and Jannach}
\institute{University of Klagenfurt, Klagenfurt, Austria \\
\email{faisal.shehzad@aau.at, dietmar.jannach@aau.at}
}

\maketitle
\begin{abstract}
In session-based recommendation settings, a recommender system 
has to base its suggestions on the user interactions that are observed in an ongoing session. Since such sessions can consist of only a small set of interactions, various approaches based on Graph Neural Networks (GNN) were recently proposed, as they allow us to integrate various types of side information about the items in a natural way.
Unfortunately,  a variety of evaluation settings are used in the
literature, e.g., in terms of protocols, metrics and baselines, making it difficult to assess what represents the state of the art.
In this work, we present the results of an
evaluation of eight recent GNN-based approaches that were
published in high-quality outlets.
For a fair comparison, all models are systematically tuned and tested under identical conditions using three common datasets. We furthermore include k-nearest-neighbor and sequential rules-based models as baselines, as such models have previously exhibited competitive performance results for similar settings. To our surprise, the evaluation showed that the simple models outperform \emph{all} recent GNN models in terms of the Mean Reciprocal Rank, which we used as an optimization criterion, and were only outperformed in three cases in terms of the Hit Rate. Additional analyses furthermore reveal that several other factors that are often not deeply discussed in papers, e.g., random seeds, can markedly impact the performance of GNN-based models. Our results therefore \emph{(a)} point to continuing issues in the community in terms of research methodology and \emph{(b)} indicate that there is ample room for improvement in session-based recommendation.
\end{abstract}


\section{Introduction}
\label{sec:introduction}
Recommender systems play a critical role in modern platforms by recommending personalized content to users according to their past preferences. Conventional recommender systems e.g., in particular ones based on collaborative filtering, rely heavily on rich user profiles and long-term historical interactions. Such systems may however perform poorly in
real-world applications, where users interact with the service without logging in~(\cite{wang2019collaborative, pang2022heterogeneous}). Consequently, session-based recommenders (SBRS), which recommend a next item solely based on an active session, received immense attention from industry and academia in recent years~\cite{Jannach2021SessionRSHB}.

From a technical point of view, most recent session-based algorithms are based on neural methods. A landmark work in this area is GRU4Rec~\cite{hidasi2015session}, which is often used as a baseline to benchmark newly developed models. One main characteristic of GRU4Rec and many subsequent models is that they are designed to operate solely on user--item interaction data. In the more recent literature, Graph Neural Networks (GNN) however received increased attention in the context of SBRS, because they allow us to consider various types of heterogeneous information in the learning process in a natural way. Examples of such recent works published in top-tier conferences in the last three years include~\cite{xu2022category,yu2020tagnn,lai2022attribute, xia2021self, zhang2021fusion, zhang2021graph, wang2020global}.

Quite surprisingly, for the category of neural SBRS models that operate solely on user--item interactions like GRU4Rec, various studies have shown that simpler methods, e.g., based on nearest-neighbor techniques, can lead to competitive accuracy results and in many cases even outperform the more complex neural models~(\cite{JannachLudewig2017c,garg2019sequence,kouki2020lab,ludewig2021empirical}). Similar observations were made for traditional \emph{top-n} recommendation tasks~(\cite{ferraridacremaetal2019,rendle2019difficulty,ferraridacrema2020tois, anelli2022top, Rendle2022Krichene}), where the latest neural models were found to be outperformed by longer-existing approaches, e.g., based on matrix factorization.\footnote{Such phenomena were also found outside the area of recommender systems, e.g., in information retrieval or time series forecasting (\cite{lin2019neural,makridakis2018statistical}).} Various factors that can contribute to this phenomenon of `phantom progress' were discussed in the literature~\cite{cremonesi2021progress}. Besides the issue that the baseline algorithms in many cases may not have been properly tuned in the reported experiments~\cite{shehzad2023everyone}, a central issue lies in the selection of the baseline models themselves. In many published research works, only very recent neural models are considered, and a comparison with well-tuned longer-existing models is often missing.

In this present work, we examine to what extent such phenomena can be found in the most recent literature on SBRS that are built on Graph Neural Networks. For this purpose, we benchmarked eight GNN models---all were recently published at top-tier venues---with simple approaches. All algorithms are reproduced under identical settings by incorporating the original code published by the authors\footnote{This is important, as third-party implementations may be unreliable~\cite{Hidasi2023TheEffect}.}  into the \emph{session-rec} SBRS evaluation framework \cite{ludewig2019performance}. The results of our study are again surprising. They show that under our independent evaluation, all of the analyzed GNN methods are outperformed by simple techniques, which do not even use side information, in terms of the Mean Reciprocal Rank, which was also our hyperparameter optimization criterion. Only in some situations, GNN-based models were favorable in terms of the Hit Rate. Overall, the results indicate that the problem of the inclusion of too weak baselines still exists in the recent SBRS literature.

While examining the research literature, we encountered a number of additional bad practices that may contribute to the apparently somewhat limited progress in this area. First, we find that researchers often use the same embedding size for all compared models for `fair comparison'. Since embedding sizes are however a hyperparameter to tune, such a comparison may instead be rather unfair (\cite{chamberlain1705neural, wu2020comprehensive}). Furthermore, the analyses in \cite{Zhu2020AreWeEvaluating} showed that in a substantial fraction of today's research, the proposed models are tuned on test data instead of using a held-out validation dataset, potentially leading to data leakage and overfitting issues that produce overly optimistic results \cite{cremonesi2021progress}.  Finally, as reported in \cite{picard2021torch}, even the choice of random seeds can have a non-negligible impact on the observed results.

In order to understand to what extent such bad practices may impact the performance results obtained by GNN models, we conducted a series of additional analyses using a subset of the models and datasets that were used in our main experiments. Our results show that the described factors can have a marked impact on the performance of the GNN-based models, thus potentially further contributing to a limited reliability of the reported progress in the literature.
\section{Research Methodology}

As discussed in earlier works \cite{ferraridacrema2020tois}, researchers often rely on a variety of experiment setups, using different protocols, metrics, datasets, pre-processing steps, and baselines. The primary objective of the study in this present paper is to provide an independent evaluation of recent GNN-based algorithms under identical conditions, i.e., same protocol, dataset, and metrics.

\subsection{Algorithms}

\emph{Compared GNN-based Models:}
To ensure that  our analysis is not focused on a few hand-picked examples, 
we followed a semi-systematic approach for the selection of algorithms to compare, using the following criteria: 
\begin{itemize}
  \item Publication outlets: We manually browsed important outlets for recommender system research (e.g., conferences such as RecSys or SIGIR and journals with a high impact factor), to identify recent works that propose GNN-based models for session-based recommendation.
  \item Reproducibility: We then only considered works for which the code was shared. We analyzed the shared repositories and contacted the authors if any part of the code was missing.\footnote{We considered articles to be non-reproducible if the authors did not reply after sending 
      reminders.}  Ultimately, we were able to identify eight GNN models that could be trained and evaluated using the provided code and which we integrated into the \emph{session-rec} framework.\footnote{\url{https://github.com/rn5l/session-rec}} The eight models are briefly described in Table~\ref{tab:tableGNNDes}.
\end{itemize}
\begin{table}[h!]
		\caption{Description of compared GNN models}
		\label{tab:tableGNNDes}
		\centering
		\begin{tabular}{p{3.5cm}  p{8.5cm} }
		\hline	
			Model                    & Description  \\
        \hline
            SR-GNN (AAAI '19)                        & Maybe the first work that uses a GNN model for SBRS. It constructs session graphs and uses a soft attention mechanism to aggregate information among the items \cite{wu2019session}.  \\

            TAGNN
            (SIGIR '20)                        & Uses an attention module attached to a GNN to adaptively learn the users' interests \cite{yu2020tagnn}.  \\
			
            GCE-GNN (SIGIR '20)                 & Relies on a two-level item embedding integrated with a position-aware attention module to recursively learn the position of items in a session \cite{wang2020global}.   \\

            COTREC  (CIKM '21)                   & Based on self-supervised learning with co-training to handle data sparsity  and to generate self-supervised signals for SBRS \cite{xia2021self}.  \\

            GNRWW (ICDM '21)                     & Combines two kinds of embedding techniques to retrieve local and global patterns from the sessions  \cite{zhang2021graph}.   \\

            FLCSP (J. Inf. Sci. '21)      & Combines latent category abstractions with sequential data to obtain accurate recommendations \cite{zhang2021fusion}. \\ 

            CM-HGNN (KBS '22)  & Differently to FLCSP, CM-HGNN leverages actual category information for accurate recommendations \cite{xu2022category}.  \\

            MGS (SIGIR '22)                        & Another model \cite{lai2022attribute} which leverages category information. Uses a dual refinement mechanism to move the information between item and category graphs. \\
            \hline

		\end{tabular}
\end{table}

\begin{table}[h!]
		\caption{Description of baseline algorithms}
		\label{tab:tableBlDes}
		\centering
		\begin{tabular}{p{2cm}  p{10cm} }
		\hline	
			Algorithm & Description  \\
        \hline

            SR \cite{ludewig2018evaluation}    &  Learns sequential rules of order two by considering the co-occurrence of items in a session.    \\

            STAN \cite{garg2019sequence}       & Builds on SKNN \cite{ludewig2018evaluation} and it uses three decay factors to compute the similarity of sessions and the relevance of the items.  \\

            VSTAN \cite{ludewig2021empirical}  & Combines the particularities of VSKNN \cite{ludewig2018evaluation} and STAN and uses a sequence-aware item scoring procedure and an Inverse Document Frequency approach to promote less popular items.   \\

            SFSKNN \cite{ludewig2018evaluation}     & This variant of SKNN \cite{ludewig2018evaluation} also focuses on the recency of items by considering only those items for recommendations that appear in the neighbor sessions at least once after the last item of the current session.   \\
            \hline

		\end{tabular}
\end{table}

\emph{Baselines:} We included four non-neural baseline models that were also used in previous performance comparisons in \cite{kouki2020lab} and \cite{ludewig2019performance}, where it turned out that such more simple models can sometimes be quite difficult to beat. The baselines are briefly described in Table~\ref{tab:tableBlDes}.

\subsection{Datasets and Preprocessing}

\emph{Datasets:}
We inspected the relevant literature from high-quality outlets to identify the most frequently used datasets.  While a variety of datasets is used, the most prominent ones include RSC15 (RecSys Challenge 2015), DIGI (Diginetica), and RETAIL (Retailrocket).\footnote{\url{https://www.kaggle.com/datasets/chadgostopp/recsys-challenge-2015},\\ \url{https://competitions.codalab.org/competitions/11161} \\ \url{https://www.kaggle.com/datasets/retailrocket/ecommerce-dataset}} Another important reason for choosing these datasets is that they contain category information, which is used by some GNN-based models. Table \ref{tab:statDataset} provides summary statistics for the selected datasets.

\setlength{\tabcolsep}{6pt}
\begin{table}[h!]
		\caption{Dataset statistics}
		\label{tab:statDataset}
		\centering
		\begin{tabular}{lccccc}
		\hline	
                Dataset      &  Clicks         & Items      & \# Categories & Sessions       & Avg.~sess. length  \\
        \hline
                RSC15 (1/12)       & 574,157     & 20,979    & 58        & 145,202       & 3.95  \\
                RSC15 (1/64)       & 107,242      & 13,133   & 31        & 26,464        & 4.05  \\
                DIGI               & 1,004,598   & 43,100    & 995       & 216,134       & 4.65  \\
                RETAIL             & 1,045,413   &44,540     & 944       & 304,902       & 3.42  \\
        \hline
		\end{tabular}
\end{table}
\emph{Data filtering and splitting:} We adopt common data preprocessing practices discussed from the literature~(\cite{zhang2020learning,xu2022category,liu2018stamp}). Sessions of length one, items that appear less than five times in the dataset, and items that occur in test data but are not present in the training dataset are filtered out. From a training perspective, RSC15 consists of a large number of sessions, making it difficult to systematically tune all GNN models for large dataset without having access to massive GPU resources. Since previous works~(\cite{song2019session,ludewig2018evaluation}) show that using a recent fraction of the data leads to competitive or even superior results, we use the commonly used 1/12 and 1/64 fractions of the RSC15 dataset. For data splitting, we follow the common practice as mentioned in~(\cite{wang2022spatiotemporal,wang2019collaborative}). For DIGI, the sessions of the last seven days are put aside as test data and the remaining sessions are used for hyperparameter tuning and training. For RSC15 and RETAIL, the last-day sessions are used as test data.

\emph{Discussion.} While our choices for data preprocessing align with what is common in the literature, our analysis of the relevant papers revealed that there is no agreed standard. Some works, for example, only consider items that appear in at least 30 sessions, others filter out sessions that contain more than 20 items. In yet another set of papers, data preprocessing is not mentioned at all. All of this further contributes to the difficulty of determining the state of the art.
In terms of the used datasets, we focused on three widely used ones. We note that each of the 8 considered GNN-based models listed in Table~\ref{tab:tableGNNDes} used at least one of these three datasets. Further experiments with other datasets to are left for future work.

\subsection{Evaluation Metrics and Tuning}
\emph{Metrics and Evaluation Protocol:} We rely on Mean Reciprocal Rank and Hit Rate (MRR@K, HR@K) as measures that are widely used\footnote{Accuracy metrics such as Precision, Recall, or NDCG are used as well in session-based recommendation, but these metrics are often highly correlated (\cite{DBLP:conf/recsys/ValcarceBPC18,anelli2022top}).} in the relevant literature~\cite{yuan2020future}.
Furthermore, we report two beyond-accuracy metrics, coverage and popularity. We found that different evaluation protocols are used in the original papers that proposed the eight GNN-based models, e.g., leaving out the last item of each test session. In our evaluation, we relied on the commonly used procedure of incrementally ``revealing'' the items in the session. This entails that all items but the first one in a session are part of the ground truth in the evaluation at some stage, see also (\cite{hidasi2015session,ludewig2018evaluation}).

\emph{Hyperparameter optimization:} We tune the hyperparameters of all models using the training dataset and validate them using a subset of the training dataset. As it is commonly known, the tuning process of GNN models can be time-consuming, in particular when a larger hyperparameters space is searched. Therefore, we relied on a random optimization process where the number of training rounds is 25 to 60, depending on the time complexity of the selected models~\cite{ludewig2019performance}. All selected models for the three datasets are optimized using MRR@20 as the target metric. The ranges and tuned values of the hyperparameters can be found in an online repository, where we also share all data and the code used for pre-processing, tuning, and evaluation for reproducibility.\footnote{\url{https://faisalse.github.io/SessionRecGraphFusion/}}
\section{Results}
\label{sec:results}

In this section we first report the main results of our performance comparison in Section~\ref{subsec:performance-measures} and then provide additional analyses in Section~\ref{subsec:robustness-analyses}.

\subsection{Performance measures}
\label{subsec:performance-measures}
In this section, we first report the results for the accuracy and beyond-accuracy quality measures and then briefly discuss aspects of computational complexity.

\emph{Accuracy.} Table \ref{tab:accuracyResults} shows the accuracy results for the different datasets at the common list lengths of 10 and 20. We sort the results by MRR@20, as this was our optimization target. In terms of the MRR, we find that for all datasets one of the simple methods leads to the highest accuracy values. In particular STAN and VSTAN show consistently good results across several datasets. There is however no single `winner' across datasets. For example, SR works particularly well for the larger RSC15 (1/12) dataset.\footnote{The effectiveness of SR on some datasets in terms of MRR was observed also in~\cite{ludewig2018evaluation}.} Generally, we observe that the differences between the winning simple method and the best GNN-based model are sometimes quite small. Nonetheless, it is surprising that several years after the first publication on the effectiveness of nearest-neighbor techniques in 2017 \cite{JannachLudewig2017c}, the most recent published models are not consistently outperforming these baselines.

Looking only at the GNN models, we find there is also no model that is consistently better than the others. The ranking of the GNN-based models varies across datasets and depending on the chosen metric. While this is not surprising, this observation stands in contrast to almost all published works on SBRS, where any newly proposed model is usually reported to outperform all other baselines on all datasets and metrics. Moreover, we find that some of the recent GNN-based models, even though reported to outperform the state of the art in the original papers, actually perform very poorly in our comparison, leading to MRR values that are sometimes more than 50\% lower than the best models.

In terms of the Hit Rate, we observe that VSTAN is the best model for the RETAIL dataset, and it has highly competitive performance for the RSC15 (1/12) dataset. For the DIGI and RSC15 (1/64) datasets, in contrast, several GNN-based models are better than the simpler approaches. In particular for the DIGI dataset, the margin between the best GNN model and the best simple model is quite large. This suggests that GNN models in fact can be effective and help us to achieve progress over existing models. The observed improvements at least in our experiment are limited to certain configurations and to the Hit Rate. We however recall that the Hit Rate was not the optimization goal during hyperparameter tuning.

Considering only the rankings of the GNN-based models, again no clear winner can be found. In some cases, models that use category information (in particular MGS) work quite well. In other cases, the reliance on category information does not seem to be too helpful, and other category-agnostic GNN models lead to higher accuracy.
\begin{longtable}{l c c c c}
    \caption{Accuracy results, sorted by MRR@20. Black circles in the table indicate simple baselines and empty circles indicate GNN-based models. The highest scores for each metric are printed in bold font while the second-best scores are underlined.}  \label{tab:accuracyResults} \\
    \toprule
        Metrics             & MRR@10            &  MRR@20           & HR@10             & HR@20  \\
    \midrule
\endfirsthead
    \caption[]{Accuracy results, sorted by MRR@20. (cont.).}    \\
    \toprule
     Metrics             & MRR@10            &  MRR@20           & HR@10             & HR@20  \\
     \midrule
\endhead
    \midrule
    \multicolumn{5}{r}{\footnotesize\itshape Continue on the next page}
\endfoot
    \bottomrule
\endlastfoot

    \multicolumn{5}{c}{RETAIL}   \\
    \hline
        $\bullet$ VSTAN     & \textbf{0.631}    & \textbf{0.631}    & \textbf{0.971} & \textbf{0.980} \\
        $\circ$ SR-GNN      & \underline{0.629} & \textbf{0.631}    & 0.886             & 0.931 \\
        $\bullet$ SFSKNN    & 0.599             & \underline{0.603} & \underline{0.939} &\textbf{0.980} \\
        $\circ$ CM-HGNN     & 0.562             & 0.568             & 0.812             & 0.890 \\
        $\bullet$ SR        & 0.553             & 0.560             & 0.865             & \underline{0.959} \\
        $\circ$ GNRRW       & 0.553             & 0.558             & 0.804             & 0.869 \\
        $\circ$ MGS         & 0.553             & 0.558             & 0.820             & 0.878 \\
        $\circ$ COTREC      & 0.551             & 0.556             & 0.792             & 0.861 \\
        $\bullet$ STAN      & 0.544             & 0.548             & 0.873             & 0.931 \\
        $\circ$ TAGNN       & 0.540             & 0.545             & 0.804             & 0.882 \\
        $\circ$ FLCSP       & 0.451             & 0.455             & 0.800             & 0.865 \\
        $\circ$ GCE-GNN      & 0.423             & 0.429             & 0.596             & 0.678 \\

    \hline
        \multicolumn{5}{c}{RSC15 (1/64)}   \\
    \hline
        $\bullet$ STAN      & \textbf{0.290}    & \textbf{0.296}    & 0.538             & 0.613 \\
        $\bullet$ VSTAN     & \underline{0.286} & \underline{0.289} & \underline{0.546} & 0.595 \\
        $\circ$ GCE-GNN      & 0.278             & 0.285             & 0.538             & \underline{0.633} \\
        $\circ$ CM-HGNN     & 0.278             & 0.284             & \textbf{0.575}    & \textbf{0.650} \\
        $\bullet$ SFSKNN    & 0.264             & 0.264             & 0.422             & 0.428 \\
        $\circ$ COTREC      & 0.274             & 0.279             & 0.543             & 0.616 \\
        $\circ$ GNRRW       & 0.269             & 0.276             & 0.526             & 0.618 \\
        $\circ$ MGS         & 0.264             & 0.270             & 0.543             & 0.630 \\
        $\bullet$ SR        & 0.263             & 0.266             & 0.462             & 0.506 \\
        $\circ$ SR-GNN      & 0.245             & 0.251             & 0.497             & 0.581 \\
        $\circ$ TAGNN       & 0.195             & 0.201             & 0.425             & 0.509 \\
        $\circ$ FLCSP       & 0.176             & 0.183             & 0.393             & 0.497 \\

    \hline
        \multicolumn{5}{c}{RSC15 (1/12)}   \\
    \hline
        $\bullet$ SR        & \textbf{0.338}    & \textbf{0.344}    & 0.557             & 0.627 \\
        $\circ$ GNRRW       & \underline{0.335} & \underline{0.342} & 0.581             & \textbf{0.682} \\
        $\circ$ SR-GNN      & 0.328             & 0.332             & 0.563             & 0.621 \\
        $\circ$ MGS         & 0.326             & 0.331             & 0.566             & 0.642 \\
        $\bullet$ STAN      & 0.325             & 0.330             & \underline{0.590} & 0.661 \\
        $\bullet$ VSTAN     & 0.325             & 0.330             & \textbf{0.599}    & \underline{0.673} \\
        $\bullet$ SFSKNN    & 0.321             & 0.325             & 0.550             & 0.609 \\
        $\circ$ COTREC      & 0.307             & 0.312             & 0.560             & 0.639 \\
        $\circ$ CM-HGNN      & 0.293             & 0.299             & 0.532             & 0.618 \\
        $\circ$ FLCSP       & 0.284             & 0.290             & 0.480             & 0.560 \\
        $\circ$ TAGNN       & 0.281             & 0.288             & 0.520             & 0.621 \\
        $\circ$ GCE-GNN      & 0.220             & 0.225             & 0.391             & 0.453 \\
    \hline
        \multicolumn{5}{c}{DIGI}   \\
    \hline

       $\bullet$ SFSKNN     & \textbf{0.348}    & \textbf{0.351}    & 0.559             & 0.604 \\
        $\bullet$ STAN      & \underline{0.347} & \textbf{0.351}    & 0.529             & 0.600 \\
        $\bullet$ VSTAN     & 0.342             & \underline{0.346} & 0.520             & 0.581 \\
        $\bullet$   SR      & 0.333             & 0.337             & 0.568             & 0.617 \\
        $\circ$ COTREC      & 0.330             & 0.335             & 0.555             & 0.637 \\
        $\circ$ MGS         & 0.322             & 0.329             & 0.559             & 0.656 \\
        $\circ$ SR-GNN      & 0.321             & 0.327             & \textbf{0.591}    & \textbf{0.688} \\
        $\circ$ GNRRW       & 0.318             & 0.324             & \underline{0.578} & \underline{0.667} \\
        $\circ$ CM-HGNN     & 0.310             & 0.316             & 0.561             & 0.649 \\
        $\circ$ TAGNN       & 0.279             & 0.287             & 0.544             & 0.645 \\
        $\circ$ GCE-GNN      & 0.227             & 0.236             & 0.419             & 0.553 \\
        $\circ$ FLCSP       & 0.175             & 0.184             & 0.398             & 0.525 \\

\end{longtable}

An ANOVA analysis revealed that there are significant differences (p$<$0.05) for the obtained MRR@20 values for the DIGI and RETAIL datasets across all examined models. However, the differences between the best-performing models for these datasets is too small to reach statistical significance according to a Tukey post-hoc test. Nonetheless, given these results, we can conclude that the best simple models are performing at least equally well as the neural ones.

\emph{Additional quality measures.}
In Table~\ref{tab:beyondaccuracy} we report additional beyond-accuracy metrics that are relevant in practice. Cov@20 refers to the percentage of items, which appear at least once in the top-20 recommendation lists for all test sessions. Pop@20 of an item is measured in terms of the number of times it appears in the training sessions. The metric reported here is the average (normalized) popularity of the recommended items in the top-20 recommendation lists for all test sessions. The results show that the simple models lead to high coverage values, indicating that these models consider a broader range of items in their recommendations at an aggregate level.
In terms of popularity, we find that the simple methods are at about the same level as some of the GNN-based approaches.
However, two GNN-based models, namely TAGNN, and GCE-GNN seem to have a stronger and usually undesired tendency to recommend more popular items.

\emph{Time complexity.} To illustrate the complexity of the problem, we report an example of the training time  (T-Time) and prediction time (P-Time) for the different models on the DIGI dataset using our hardware\footnote{A machine with an AMD EPYC 7H12 64-Core Processor 2600 Mhz 16 Core(s) and an NVIDIA RTX A4000 WDDM graphics card.} in Table~\ref{tab:beyondaccuracy}. On this dataset, GCE-GNN is the slowest model which takes approximately 19 hours to be trained once, and we recall that many training rounds are needed during hyperparameter tuning for each dataset and for each model. We also note that the training time may vary strongly between the GNN models across datasets, depending, e.g., on the number of items, the number of the sessions, and the length of the sessions. The kNN-based models, in contrast, do not have a training phase. The time needed for predictions for these models lies about in the range of the faster GNN-based models. However, we notice that the COTREC model is particularly slow and needs approximately 2 seconds to generate one single recommendation list, which is prohibitive in practice.

\begin{table}[!ht]
    \caption{Results for the DIGI dataset, sorted by Cov@20}
    \centering
    \label{tab:beyondaccuracy}
    \begin{tabular}{l c c r r}
    \hline
        Metrics   & Cov@20                        & Pop@20            &  T-Time (m)             & P-Time (ms)  \\
    \hline
        $\bullet$ STAN      & \textbf{0.087}      & 0.091             & \textbf{0.038}          & 14.697 \\
        $\bullet$ VSTAN     & \underline{0.086}   & 0.084             & \underline{0.045}       & 13.911 \\
        $\bullet$ SR        & \underline{0.086}   & 0.075             & 0.121                   & \textbf{3.575} \\
        $\circ$ SR-GNN      & 0.079               & 0.078             & 55.119                  & 27.468 \\
        $\circ$ COTREC      & 0.078               & 0.079             & 480.709                 & 1992.498 \\
        $\circ$ MGS         & 0.076               & 0.083             & 390.894                 & 14.635 \\
        $\circ$ TAGNN       & 0.073               & \textbf{0.102}    & 119.823                 & 6.991 \\
        $\circ$ GNRRW       & 0.073               & 0.078             & 154.657                 & 173.524 \\
        $\circ$ FLCSP       & 0.073               & 0.075             & 40.825                  & 10.393 \\
        $\circ$ CM-HGNN     & 0.072               & 0.083             & 944.725                 & 12.839 \\
        $\bullet$ SFSKNN    & 0.072               & 0.071             & 0.081                   & \underline{5.505} \\
        $\circ$ GCE-GNN     & 0.065               & \underline{0.097} & 1,153.861               & 503.492 \\
    \hline
    \end{tabular}
\end{table}

\subsection{Additional Robustness Analyses for GNN-based Models}
\label{subsec:robustness-analyses}
As mentioned earlier, a number of factors can endanger the robustness and reliability of
accuracy results reported in research papers, including the choice of embedding sizes and random seeds, or the practice of tuning on test data. When we were screening the literature for recent GNN-based models, we identified 34 relevant articles in top-level outlets.\footnote{The list of considered papers can be found in the online material.} An analysis of these papers leads to a number of interesting observations.

\emph{Embedding size.} We found that the authors frequently fix the embedding size and only tune the other hyperparameters such as learning rate or dropout rate. In several cases, we also observed that authors chose the embedding size of the baseline models from the original papers which may have been optimal for different datasets. However, they tune the embedding size of their own proposed models. Such a comparison, in which only one of the models is extensively tuned, certainly cannot lead to any reliable insights, as the embedding size in fact is a crucial hyperparameter (\cite{chamberlain1705neural, wu2020comprehensive, hamilton2017representation}).

To analyze the potential extent of the problem, we conducted experiments in which we varied the embedding size of the models while keeping the other hyperparameters constant (at their optimal values, as determined in the previous experiment). Since some GNN-based models are computationally expensive, we focused on five training-efficient models and on two datasets.

As a representative of a smaller dataset, we chose the RSC15 (1/64) dataset and conducted the experiments by using 15 different values for embedding sizes ranging from 16 to 500. We used the DIGI dataset as an example for a larger dataset, and the experiments are conducted using 10 different values for embedding sizes ranging from 16 to 250. In this case, the maximum value of the embedding size is 250, as we experienced memory problems with larger values.

The outcomes of these experiments are provided in Table \ref{tab:embeddingSize}.\footnote{Note: the MRR@20 values of some GNN models are not the same as in Table 4 because we decreased the values of batch sizes to avoid memory issues, which in turn affects the accuracy of the models.} We can observe that some models, depending on the embedding size, can lead to largely different results in terms of the MRR@20. This confirms the importance of carefully tuning this central hyperparameter. As the results show, the differences between the best and worst MRR values can be more than 40\% in some cases, simply by choosing a bad value for the embedding size. For some of the models, the difference between the minimum and maximum MRR values is less pronounced, but still in a range where researchers would report a substantial improvement over the state of the art when comparing models. We note that no clear pattern was found across the models that larger embedding sizes are consistently better than smaller ones. Furthermore, the effect size ($\eta^{2}$) is measured using the ANOVA test and found that its values for the DIGI and RSC15 datasets are 85.53\% and 91.39\% respectively, indicating the high variance between the mean MRR values of the GNN models.

\begin{table}[!ht]
    \caption{Results with varying embedding sizes, sorted based on difference}
    \centering
    \label{tab:embeddingSize}
    \begin{tabular}{l |c| c |c|c}
    \hline
        Models      & MRR@20 mean $\pm$ std         & Max MRR@20       & Min MRR@20 & Diff \\
    \hline
        \multicolumn{4}{c}{RSC15 (1/64)}   \\
    \hline

        GNRWW       & 0.237  $\pm$  0.025           & 0.291     & 0.206 &  0.085 \\
        TAGAN       & 0.195  $\pm$  0.021           & 0.226     & 0.143 &  0.083 \\
        FLCSP       & 0.160  $\pm$  0.014           & 0.180     & 0.135 &  0.045 \\
        SR-GNN      & 0.238  $\pm$  0.012           & 0.259     & 0.218 &  0.041\\
        COTREC      & 0.277  $\pm$  0.003           & 0.284     & 0.269 &  0.015 \\

    \hline
        \multicolumn{4}{c}{DIGI}   \\
    \hline
    TAGAN           & 0.247   $\pm$   0.033            & 0.294     & 0.164 & 0.130  \\
    COTREC          & 0.335   $\pm$    0.010           & 0.343     & 0.309 & 0.034 \\
    FLCSP           & 0.206   $\pm$    0.011           & 0.224     & 0.191 & 0.033 \\
    SR-GNN          & 0.313   $\pm$    0.008           & 0.324     & 0.296 & 0.028  \\
    GNRWW           & 0.336   $\pm$    0.005           & 0.346     & 0.327 & 0.019  \\
    \hline
    \end{tabular}
\end{table}

\begin{figure}
    \centering
     \begin{subfigure}[b]{0.35\textwidth}
         \centering
         \includegraphics[width=\textwidth]{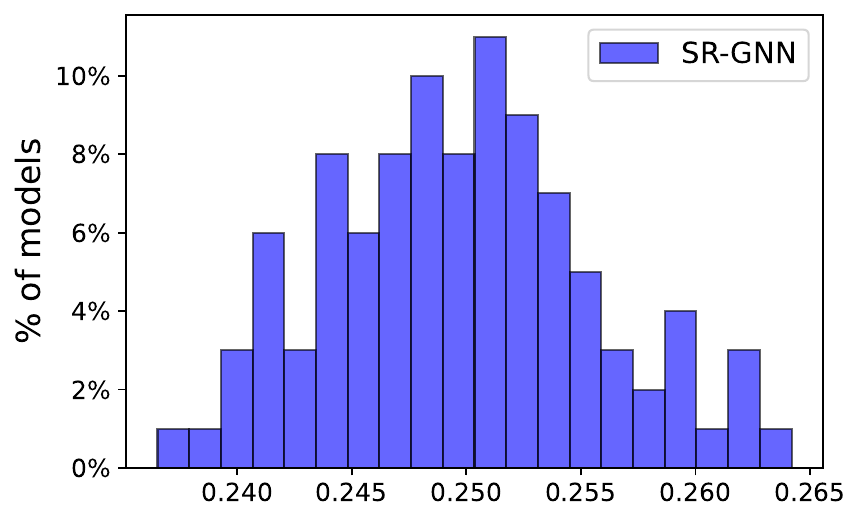}
     \end{subfigure}
     \begin{subfigure}[b]{0.35\textwidth}
         \centering
         \includegraphics[width=\textwidth]{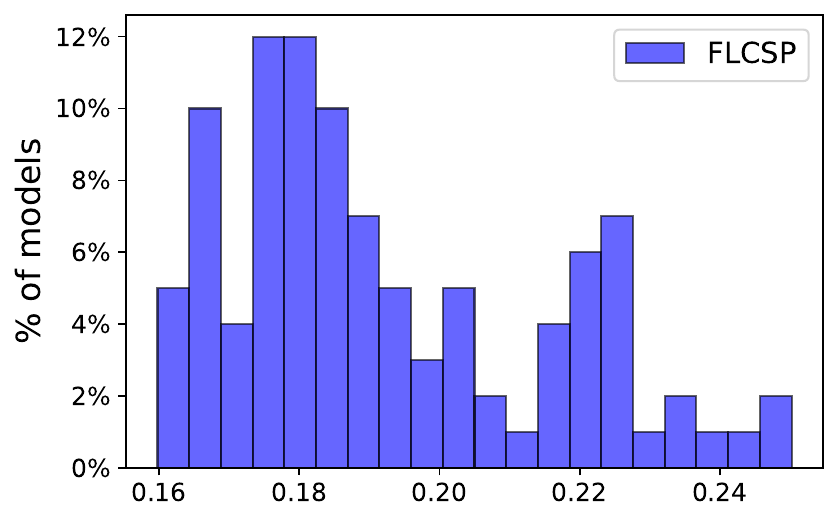}
     \end{subfigure}
        \caption{Distribution of MRR@20 values for different random seeds (RSC15)}
        \label{fig:seedsfig}
\end{figure}

\emph{Random seeds.} The random seed used in an experiment is certainly not a hyperparameter to tune. Still, its choice can have notable impact on accuracy results (\cite{wegmeth2023effect, picard2021torch}). 
We therefore conducted experiments to analyze to what extent this is true also for some of the examined GNN-based models. In these experiments, we selected random seed values between 100 to 10,000,000 by using a random function. In the shared repositories of the GNN models, random seed values between 2000 to 3000 can often been found, but there is no general pattern. In our experiments, we deliberately picked a larger range to examine if some more uncommon values (outliers) could have an unexpected effect. Again, we considered the RSC15 (1/64) and the DIGI dataset. We generated 100 random seed values for the RSC15 dataset, and 35 values for the larger DIGI dataset. The hyperparameters of the examined models were again left at their optimum values.
The obtained results for MRR@20 when using different random seeds are shown in Table~\ref{tab:randomseeds}, again sorted by the difference between the minimum and the maximum observed value. Again, we can see that for some models and datasets, the obtained values using the same set of optimized hyperparameters vary strongly. For the RSC15 dataset and the FLCSP method, for example, the obtained accuracy results can be 30\% lower than the best value simply because of a unlucky choice of the random seed value.\footnote{In this case, the random seed value leading to the worst MRR value was around 7,000,000. All detailed results can be found online.} Clearly, the differences are not as extreme for all models and datasets and may be partly due to the large range of values that we explored. The distribution of the accuracy values Figure~\ref{fig:seedsfig} for the RSC15 (1/64) dataset. The detailed data of the entire experiment can be found in the online material.

Comparing the results of Table~\ref{tab:randomseeds} and the results obtained when using the random seed values used in the original code shared by the authors (Table~\ref{tab:accuracyResults}), we observe that often higher accuracy values can be achieved when exploring a large number of random seed values. For example, the optimal MRR@20 value for the FLCSP method on the RSC15 (1/64) dataset is 0.183. From the first row in Table~\ref{tab:randomseeds} we can see that the worst MRR value was at 0.159 and the best one at 0.250, which can be seen as a substantial improvement, which is only obtained by changing the random seed. A similar observation can be made in other situations as well, e.g., considering the last row of Table~\ref{tab:randomseeds}, where the best MRR@20 value of GNRWW is at 0.352 
for the DIGI dataset, whereas the best value in Table~\ref{tab:accuracyResults} was at 0.324. Note, however that in some cases the best MRR values obtained by exploring many (somewhat unusual) seed values are lower than when using more common choices as can be found in the shared code repositories, which means that these common ranges should be explored as well. Furthermore, $\eta^{2}$ is measured and found to be 97.79\% and 89.56\% for the DIGI and RSC15 datasets respectively, indicating the high variance between the mean MRR values of the GNN models.

\begin{table}[!ht]
    \caption{Experiments with random seeds, sorted based on difference}
    \centering
    \label{tab:randomseeds}
    \begin{tabular}{l |c| c |c | c}
    \hline
        Models      & MRR@20 mean $\pm$ std         & Max  MRR@20   & Min MRR@20           & Diff  \\
    \hline
        \multicolumn{4}{c}{RSC15 (1/64)}   \\
    \hline
    FLCSP           & 0.192  $\pm$  0.022           & 0.250             & 0.159                     & 0.091   \\
    TAGNN           & 0.205  $\pm$  0.011           & 0.232             & 0.175                     & 0.057   \\
    GNRWW           & 0.273  $\pm$  0.009           & 0.293             & 0.245                     & 0.048   \\
    SR-GNN          & 0.249  $\pm$  0.005           & 0.264             & 0.236                     & 0.028   \\
    COTREC          & 0.283  $\pm$  0.003           & 0.290             & 0.275                     & 0.015   \\
    \hline
        \multicolumn{4}{c}{DIGI}   \\
    \hline
    TAGNN          & 0.256    $\pm$ 0.015           & 0.281             & 0.228                     & 0.052   \\
    FLCSP           & 0.203   $\pm$ 0.009           & 0.217             & 0.172                     & 0.045   \\
    SR-GNN          & 0.322   $\pm$ 0.009           & 0.342             & 0.307                     & 0.035   \\
    GNRWW           & 0.340   $\pm$ 0.006           & 0.352             & 0.327                     & 0.025   \\
    COTREC          & 0.337   $\pm$ 0.003           & 0.343             & 0.329                     & 0.014   \\
    \hline
    \end{tabular}
\end{table}

\emph{Tuning on test data.} In a last experiment, we repeated the main experiment from Section~\ref{sec:results}, but this time tuned the hyperparameters of four training-efficient GNN models on the test set, which is unfortunately not an uncommon practice according to \cite{Zhu2020AreWeEvaluating}. The results show that depending on the dataset an accuracy improvement by 1.5 to about 2.0\% in terms of MRR@20 across models can be achieved. The strongest individual ``improvement'' was at 5.6\%, which confirms the observation from \cite{cremonesi2021progress} that such evaluation practice leads to overly optimistic assessments of the accuracy of a model. The detailed results of the experiment can be found in the online material.
\section{Conclusion}
\label{sec:conclusion}
Our systematic analysis of eight recent GNN-based models for session-based recommendation reveal that methodological problems that were identified several years ago still exist in this research area. Specifically, researchers tend to rely on experimental evaluations which mostly consider other recent neural models as baselines, but do not include simpler, yet often well-performing baselines in their comparisons. Considering again the 34 recent papers from top-tier outlets that we analyzed in our study, we find that none of them considered models like SFSKNN or VSTAN as baselines. Instead, trivial methods like popularity based ones or the sometimes poorly performing \emph{item-KNN} baseline from \cite{hidasi2015session} are considered. Only one paper \cite{kouki2020lab} considered several non-neural baselines, but this was a particular study which compared the offline and online performance of various models based on the \emph{session-rec} framework.

In sum, this leads to some worries regarding the true progress that is achieved in the field, in particular when we combine this finding with other potentially problematic factors that can influence the outcomes reported in research papers, as discussed in the additional analyses in this paper. On the more positive side, our analysis suggests that there is ample room for future improvements of session-based recommendation models. In particular, not too many works exist yet that consider the various types of side information, e.g., item category, price, that are available in some datasets. We believe that approaches based on Graph Neural Networks are indeed promising for such settings with rich data, even though the methods examined in our present study do not seem to exploit their full potential yet.

In terms of future directions, we observe that various recent approaches for session-based recommendation are based on self-attention and the transformer architecture, e.g., (\cite{Fan2021Lighter,Hou2022Core}). A performance comparison of such models is part of our future work. In a recent study, it was found that \emph{sequential} transformer-based models like BERT4Rec~\cite{Sun:19} indeed seem to be outperforming simple models with a substantial gap, at least for larger datasets. An up-to-date performance comparison of \emph{session-based} models based on attention and transformers is however still missing.

\bibliographystyle{splncs04}
\bibliography{sample-base}
\end{document}